\documentclass[aps,prb,twocolumn,floatfix,showpacs]{revtex4}

\usepackage{epsfig}
\usepackage{amsmath}

\usepackage{graphicx}
\usepackage{amsfonts}
\usepackage{amssymb}

\newcommand{\be}{\begin{equation}}
\newcommand{\ee}{\end{equation}}
\newcommand{\bea}{\begin{eqnarray}}
\newcommand{\eea}{\end{eqnarray}}
\newcommand{\bm}[1]{\mathbf{#1}}

\newcommand{\ra}{\rangle}
\newcommand{\me}[3]{\langle #1 | #2 | #3 \rangle}

\begin{document}

\title{The physical picture behind the oscillating sign of drag in high Landau levels}

\author{Rafi \surname{Bistritzer}}
\author{Ady \surname{Stern}}
\affiliation{Department of Condensed Matter Physics, The Weizmann
Institute of Science, Rehovot, 76100, Israel.}

\begin{abstract}

We consider the oscillating sign of the drag resistivity and its
anomalous temperature dependence discovered experimentally in a
bi-layer system in the regime of the integer quantum Hall effect.
We attribute the oscillating sign to the effect of disorder on the
relation between an adiabatic momentum transfer to an electron and
the displacement of its position. While in the absence of any
Landau level mixing a momentum transfer $\hbar \bf q$ implies a
displacement of $ql_H^2$ (with $l_H$ being the magnetic length),
Landau level mixing induced by short range disorder adds a potentially large
displacement that depends on the electron's energy, with the sign
being odd with respect to the distance of that energy from the
center of the Landau level. We show how the oscillating sign of
drag disappears when the disorder is smooth and when the
electronic states are localized.

\end{abstract}

\maketitle

Coulomb drag in bi-layer systems at high Landau levels has been an
active area of research in the last few years, both theoretically
and experimentally. This research has been initiated by the
puzzling results obtained in a series of experiments
\cite{Feng,Lok,Muraki}, where the longitudinal drag resistivity
$\rho^D$ was measured in the presence of strong magnetic fields
and large Landau level filling factors.

In a typical drag experiment \cite{Gramila}, a current $j^{(a)}$
is driven through one layer, the active layer, while no current is
permitted to flow in the other layer, the passive layer. The
longitudinal drag resistivity is defined by
\be
E^{(p)}= -\rho^D j^{(a)}_y,              \label{rho_D_definition} \ee where
$E^{(p)}$ is the component of the induced electric field in the
passive layer that is parallel to $j^{(a)}$.

Experimentally, two anomalous features of $\rho^D$ at low
temperatures were discovered. The first is a sign that oscillates when the density difference between the two layers is
varied: when the Landau level filling factors of the two layers differ by an
even number $\rho^D$ is positive, i.e., has the same sign as at
zero magnetic field. In contrast, it becomes negative when
the filling factors of the two layers differ by an odd number. The
second anomaly is the non monotonous dependence of $\rho^D$ on the temperature $T$.
For high temperatures, $\rho^D$ has a quadratic dependence on
$T$, similar to that observed at zero magnetic field.
However, as temperature is decreased to low enough values there is
a sudden increase of $|\rho^D|$. As the temperature is further
decreased $|\rho^D|$ decreases, finally reaching zero in the limit
of $T\rightarrow 0$. Both anomalous features occur at the same
range of temperatures.

The experimental observations were theoretically addressed in
several works \cite{VonOppen,Bonsager,Khaetskii,Gornyi}. An
important step towards their understanding has been carried out by
Gornyi et al.\cite{Gornyi}, who used the Self Consistent Born Approximation
(SCBA), and found the sign of $\rho^D$ to oscillate
as the densities of the layers are varied.  The non-monotonous
temperature dependence of $\rho^D$ was obtained as well, albeit its
Arrhenius functional form at low temperature was missed. Although
the SCBA calculation is generally consistent with the experimental
observation, the physical picture behind its results is obscure.

In this work we study $\rho^D$ in a way that makes this physical picture transparent. We show
that the oscillating sign of $\rho^D$ originates from the effect of disorder on
the relation between an adiabatic momentum transfer to an electron
and the displacement of its position. Coulomb drag stems from scattering processes in which
an energy $\omega\lesssim T$ and a momentum $q$ are transferred from one layer to another
(we set $\hbar=1$ throughout this paper). In the limit of low temperature, $\omega$ is very small,
and the transfer of momentum is adiabatic. When disorder is weak enough to allow any Landau
level mixing to be neglected, this momentum transfer  implies a
displacement of $ql_H^2$.
We find that with the inclusion of lowest order Landau level mixing, the combined effect of a
short range disorder and
the rapidly oscillating wave functions characteristic of high Landau levels
results in \emph{an additional displacement}. The sign of this displacement depends
on the electron's energy relative to the center of its Landau level, and its magnitude becomes
large in the limit of a high Landau level. The electron's energy is typically the chemical potential,
and thus $\rho^D$ develops oscillations with respect to the variation of the chemical potential
in either one of the layers.
Furthermore, we show that the non-monotonic temperature dependence originates from the strong
oscillations in the density of states characteristic of well separated Landau levels.

We consider the weak
coupling regime \cite{Zheng, Kamenev,Flensberg} in which
\small
    \be
    \rho^D =  \frac{\rho^{(p)}_{yx} \rho^{(a)}_{xy}}{8T}
\int \frac{d \omega}{2 \pi} \sum_{\bm{q}} |U_{\bm{q}\omega}|^2
\frac{\Gamma_x^{(p)}(\bm{q},\omega,B)
\Gamma_x^{(a)}(\bm{q},\omega,-B)}{\sinh^2(\omega/2T)},
\label{sigma_D_xx}
    \ee \normalsize
where ${\rho}^{(p)}_{yx}$ and ${\rho}^{(a)}_{xy}$ are respectively
the Hall resistivities of the passive and active layers, $U$ is
the screened inter-layer interaction and $B$ is the magnetic
field. Most importantly, $\bm{\Gamma}$ is the rectification
function relating a scalar potential $V_{\bm{q}\omega}$ of wave
vector $\bf q$ and frequency $\omega$ to the $DC$ current it
creates in a second order response:
\be \bm{j}_{\scriptscriptstyle{DC}} = \bm{\Gamma}_{\bm{q}\omega}  | _{\bm{q}\omega}|^2.         \label{Gamma_jDC}
\ee

The function $\bm{\Gamma}$ is a vector, with a longitudinal
component $\Gamma_{||}$ parallel to the wave vector $\bf q$ and a
transverse component $\Gamma_{\perp}$ perpendicular to $\bf q$.
For symmetry reasons, $\bm{\Gamma}$ satisfies \bea
\Gamma_{\parallel}({\bf q},\omega,B)&=& -\Gamma_{\parallel}({-\bf
q},\omega,-B)
\label{Gamma_parallel_symmetry} \\
\Gamma_{\perp}({\bf q},\omega,B)&=& \Gamma_{\perp}({-\bf
q},\omega,-B)
\label{Gamma_perp_symmetry} \\
\bm{\Gamma}(-\bm{q},\omega,B) &=& -\bm{\Gamma(\bm{q},\omega,B}).
\label{Gamma_mq_symmetry} \eea Eq. (\ref{Gamma_mq_symmetry}) is
valid provided $\bm{\Gamma}$ is averaged over disorder.

Given Eqs.(\ref{sigma_D_xx}--\ref{Gamma_mq_symmetry}) and the
experimental observation of positive $\rho^D$ for identical
layers, it is clear that the drag resistivity
is dominated by the contribution of $\Gamma_{\perp}$. Thus,
the oscillations of the sign of $\rho^D$ and its $T$
dependence should originate from the dependence of
$\Gamma_{\perp}$ on $\omega$, $\bm{q}$ and the Landau level
filling factor (or, equivalently, the chemical potential). In particular, the density
dependence of the sign of $\rho^D$ must originate from a dependence of the sign of
$\Gamma_\perp$ on the chemical potential\cite{VonOppen,Gornyi}.

In view of Eq.(\ref{Gamma_perp_symmetry}) we find it
more convenient to define the Hall rectification function
\be
\Gamma_H(\bm{q},\omega) \equiv \frac{1}{2} \bigg(
\Gamma_{\perp}({\bf q},\omega,B) + \Gamma_\perp(-{\bf q},\omega,-B) \bigg).
\label{Gamma_H_definition}
\ee

We consider a model of non--interacting electrons subjected to a
magnetic field $B$ and a disorder potential $V_d$. The single
layer Hamiltonian is then, \be H = \sum_j \left\{
\frac{1}{2m}\big( \bm{p}_j - \bm{A}(\bm{r}_j) \big)^2 +
V_d(\bm{r}_j) \right\}. \ee This Hamiltonian can, in principle,
be diagonalized by single particle states with energies $\{
\epsilon_\alpha \}$ and wave functions $\{ \Psi_\alpha \}$. Hence,
the unperturbed Green functions obtain a simple single
particle form, and in the basis of exact single
particle states $\Gamma_H$ is given by the triangle
diagrams\cite{Kamenev,Flensberg}.

As we show below, within this model,
\be
\Gamma_H(\bm{q},\omega) = \sum_{\beta \gamma} \Delta x_{\beta\gamma} T_{\beta\gamma}.
\label{Gamma_H_result}
\ee
where
\be T_{\beta\gamma} \equiv   2 \pi |(\rho_{\bm{q}})_{\gamma \beta}|^2
\left[ n_{\mbox{\tiny F}}(\epsilon_\gamma) - n_{\mbox{\tiny F}}(\epsilon_\beta) \right]
\delta(\epsilon_\beta - \omega - \epsilon_\gamma)
\label{T_beta_gamma}
\ee
and
    \be \Delta x_{\beta\gamma} \equiv (x)_{\beta\beta} -
(x)_{\gamma\gamma}. \label{Delta_x}
    \ee
Hereafter the $\hat{\bm{y}}$ direction is set parallel to $\bm{q}$,
$\rho_{\bm{q}}$ is the density operator, $n_{\mbox{\tiny F}}$ is the Fermi
distribution function, and $(x)_{\beta\beta}$ is the expectation
value of the position $x$ with respect to the single particle
state $|\beta\rangle$. We discuss below the way
Eq.(\ref{Gamma_H_result}) is to be understood for extended states,
for which this expectation value may be ill defined.

Eq.(\ref{Gamma_H_result}) gives a very simple physical explanation
to the Hall rectification current $\Gamma_H({\bf
q},\omega)|V_{\bm{q}\omega}|^2$. The application of a time and
space dependent potential $V_{\bm{q}\omega}$ induces transitions
between single particle states. The net rate for transitions
between the states $\gamma$ and $\beta$ is given by Fermi's golden
rule to be $T_{\beta\gamma}|V_{\bm{q}\omega}|^2$. A transition
from a state $\gamma$ to a state $\beta$ involves also a
translation of the electron's position given by $\Delta
x_{\beta\gamma}$ , and thus induces a current. The total current
involves a sum over all possible transitions, and hence
Eq.(\ref{Gamma_H_result}).

Furthermore, Eq.(\ref{Gamma_H_result}) clarifies that the sign of
$\Gamma_H$ is determined by the preferred direction of the
induced transitions. Put differently, to determine that sign one
should identify the states $\beta$ and $\gamma$ that make the most
significant contribution to $T_{\beta\gamma}$, and determine the
sign of the displacement $\Delta x_{\beta\gamma}$ associated with their transition.

With the imaginary part of the polarization operator being ${\rm
Im}\Pi=\sum_{\beta\gamma}T_{\beta\gamma}$ we find it instructive
below to express $\Gamma_H$ in terms of ${\rm Im}\Pi$. One can view the ratio
between them as the characteristic displacement for a
particular ${\bf q},\omega$ and chemical potential.

In our discussion we are guided by the experimental conditions to focus on well separated
Landau levels, large filling factors and $q \ll k_{\mbox{\tiny F}} $, with $k_{\mbox{\tiny F}}$
being the Fermi momentum. We denote the uppermost filled Landau level by
$n$, and consider the limit of low temperatures. We also find it convenient to measure all
energies with respect to $(n+1/2)\omega_c$, the center of the $n$'th Landau level.

We first discuss the expectation value $(x)_{\beta\beta}$.
The single particle states $|\beta\rangle$ may be written as
$|\beta_0\rangle+|\beta_1\rangle$, where
$|\beta_0\rangle$ is composed of states of the $n$'th Landau
level, and  $|\beta_1\rangle$ is
composed of states of other Landau levels. To leading order in the
disorder potential, $|\beta_0\rangle =\sum_{k} C_{ k}^\beta | n k \ra$, where $\{ | nk \ra \}$ are the $n$'th Landau level wave
functions of a clean system in the Landau gauge, $\bm{A} = B(0,x)$.
The coefficients $C_k^\beta$  and the leading order approximation to the energies $\epsilon_\beta$ are found within first order
degenerate perturbation theory to satisfy,
\be
\sum_p \me{n k}{V_d}{n p} C_{p}^\beta =  \epsilon_\beta C_{ k}^\beta.
\label{iso_energetic_diagonalization}
\ee
The Landau level mixing part, $|\beta_1\rangle$, is
calculated by means of first order non-degenerate perturbation
theory and is of first order in the ratio $V_d/\omega_c$. Within that order,
\bea
(x)_{\beta\beta}=\langle\beta_0|x_0|\beta_0\rangle
-i\omega_c^{-2}\langle\beta_0|[v_x,V_d]|\beta_0\rangle.
\label{shift}
\eea
where we use the decomposition
    \be
    \bm{r} = \bm{r_0} - \bm{\hat{z}} \times
\frac{\bm{v}}{\omega_c} \label{decomposition}
    \ee
of the position
operator $\bm r$, with $\bm{r_0}$ the guiding center
coordinate and $\bm v$ the velocity operator. In Eq. (\ref{shift})
the first term is the expectation value of the guiding center coordinate $x_0$, and the second
term is the expectation value of $v_y/\omega_c$, which becomes
non-zero only due to the Landau level mixing induced by disorder.
Note that $[v_x,V_d(x)]\propto
\partial_x V_d(x)$. The shift of the position induced by the
disorder is proportional to the expectation value of the electric
field experienced by the electron.

We now turn to discuss the states $|\beta_0\rangle,|\gamma_0\rangle$.
Within the semiclassical approximation,
\be
\langle\gamma_0|\rho_{\bf
q}|\beta_0\rangle=J_0(qR_c)\langle\gamma_0|e^{-iqy_0}|\beta_0\rangle
\label{semiclass}
\ee
with $R_c$ being the cyclotron radius of the
$n$'th Landau level. Thus, in order to analyze the most
significant transitions, we consider two states: the first is a
single particle eigenstate $|\beta_0\rangle$ with an energy
$\epsilon_\beta$, in the vicinity of the chemical potential $\mu$. The second state, $|\gamma_0\rangle\equiv e^{-iqy_0}|\beta_0\rangle$, is a state defined in such a way that the potential $V({\bf q},\omega)$ couples it most effectively to $|\beta_0\rangle$: its
momentum is shifted by $q$ relative to that of
$|\beta_0\rangle$, and it is projected to the $n$'th Landau level. Furthermore, focusing on the low temperature limit of
Eqs. (\ref{sigma_D_xx},\ref{T_beta_gamma}) we look for those cases
for which the expectation value of the energy of $|\gamma_0\rangle$
equals $\epsilon_\beta$, and find the implications of this condition on $\Delta x_{\beta\gamma}$.

When the chemical potential is in a region of localized states, $|\beta_0\rangle$ and $|\gamma_0\rangle$
are both localized, and
the expectation value of the velocity, the second term in
(\ref{shift}), vanishes. Therefore,
    \be
\Delta x^{(loc)}_{\beta\gamma} = (x_0)_{\beta_0\beta_0} -
(e^{iqy_0}x_0e^{-iqy_0})_{\beta_0\beta_0} = ql_H^2,
\label{Delta_x_localized}
\ee
and
\be
\Gamma_H^{(loc)} = 2ql_H^2 {\rm Im} \Pi_{\bm{q}\omega}.
\label{Gamma_H_localized}
\ee
The sign of  $\Gamma_H$ is then independent of $\mu$.

When the states $|\beta_0 \rangle,|\gamma_0 \rangle$ are extended the
second term in (\ref{shift}) no longer vanishes and
\be
\Delta x^{(ext)}_{\beta\gamma} = ql_H^2 + \frac{1}{\omega_c}
\left( v_\beta(q) -  v_\beta(0) \right),
\label{Delta_x_extended}
\ee
where $v_\beta(q) \equiv \partial_q \epsilon(\beta,q)$ with
\be
\epsilon(\beta,q) \equiv \int \bm{dr} |\Psi_{n
\beta}(x-ql_H^2,y)|^2 V_d(x,y)
\label{epsilon_beta_q_definition}
\ee
being the energy of the state $e^{-iqy_0}|\beta_0\ra$.
For a slowly varying potential changing on a scale of $\xi \gtrsim R_c$ the
second term of Eq.(\ref{Delta_x_extended})
is of the order of $ql_H^2 \frac{V_d l_H^2}{\omega_c \xi^2}$ which is negligible compared to $ql_H^2$.
When that happens relation (\ref{Gamma_H_localized}) between $\Gamma_H$ and ${\rm Im}\Pi$
still holds, although ${\rm Im}\Pi$ now reflects the extended nature of the single particle states.

The opposite extreme is exemplified by
\be
V_d(\bm{r}) = \sum_j \delta(\bm{r} - \bm{r_j}) U_j
\label{point_like_disorder}
\ee
describing point like impurities. We assume the impurities' positions $\{{\bf r}_j\}$ to be random, and their strengths $U_j$ to be symmetric around zero.
For the simplicity of the calculations, we use a rather crude approximation for the $|nk\ra$ states,
\be
|n k \ra  =  \frac{e^{iky}}{\sqrt{R_c L_y}} \cos(k_{\mbox{\tiny F}}(x-kl_H^2))\Theta_{R_c}(x-kl_H^2)
\label{approximated_wf}
\ee
where $L_y$ is the system's length in the $\bm{\hat{y}}$
direction and $\Theta_R(x)$ is a smoothed step function that equals one for $x^2 \lesssim R^2$
and zero otherwise.
For any fixed $x$, Eq. (\ref{approximated_wf}) becomes exact in the limit $n\rightarrow\infty$. Within the approximation (\ref{approximated_wf}), the Bessel function in (\ref{semiclass}) is replaced by a cosine.
It follows from Eqs.(\ref{point_like_disorder},\ref{approximated_wf}) that
\small
\bea
\epsilon(\beta,q) &=& (R_cL_y)^{-1} \sum_{pp'j} C_{p}^{\beta*} C_{p'}^{\beta} U_j
e^{-i(p-p')y_j} \cos(\theta_{jp}+qR_c)  \nonumber \\
&& \cos(\theta_{jp'}+qR_c) \Theta_{R_c}(x_j-pl_H^2) \Theta_{R_c}(x_j-p'l_H^2),
\label{epsilon_beta_q}
\eea
\normalsize
where $\theta_{jp}=k_{\mbox{\tiny F}}(x_j-pl_H^2)$. Since $q \ll k_{\mbox{\tiny F}}$ we neglected
the change in the argument of the $\Theta$ functions .
Making the assumption that for all $y$, the disorder potential
averages to zero when integrated over $x$, i.e.,  $\int dx
V_d(x,y)=0$, we obtain from Eq.(\ref{epsilon_beta_q}) a relation
between the velocity $v_\beta(0)$ and the energies
$\epsilon_\beta$ and $\epsilon_\gamma$:
    \be \epsilon_{\gamma} =
\cos(2qR_c)\epsilon_{\beta} - \frac{1}{2R_c}v_\beta(0)\sin(2qR_c).
\label{epsilon_beta_of_epsilon_gamma}
    \ee
Being interested in the low frequency limit of
(\ref{T_beta_gamma}), we look for states $|\beta_0\rangle$ for
which $\epsilon_\beta=\epsilon_\gamma$. For these states,
\be
v_\beta(0) = -2R_c \epsilon_\beta \tan(qR_c). \label{v_beta0_2}
\ee

At low temperatures $\epsilon_\beta \approx \mu$ and hence the
sign of $v_\beta(0)$ depends on the position of $\mu$ relative to
the center of the Landau level. By interchanging the role of
states $\beta_0$ and $\gamma_0$ we find $v_\beta(q) =
-v_\beta(0)$. Thus, \be \Gamma_H^{(ext)}\approx \left( 2ql_H^2 + 4
R_c \frac{\mu}{\omega_c} \tan(qR_c) \right) {\rm Im}
\Pi_{\bm{q}\omega}. \label{Gamma_H_extended} \ee The second term
in the brackets is the crucial disordered-induced contribution to
the displacement, whose sign changes as $\mu$ crosses the center
of the Landau level. This contribution, which stems from the
second term in (\ref{Delta_x_extended}), becomes larger than the
first term for a certain range of $q$. Although it is sub-leading
in $V_d/\omega_c$, this range is made large at high Landau levels
due to the dependence of (\ref{v_beta0_2}) on $R_c$, a dependence
that originates from the rapid oscillations of the wave functions
on the scale of the Fermi wavelength. When this range is large
enough, the change of sign of $\Gamma_H$ as a function of $\mu$
leads to the sign variations of $\rho^D$ \cite{Gornyi}. We note
that within approximation (\ref{approximated_wf}) ${\rm Im}
\Pi_{\bm{q}\omega} \propto \cos^2(qR_c)$ so that
$\Gamma_H^{(ext)}$ is finite for all $qR_c$.

Eq.(\ref{Gamma_H_extended}) is similar to the SCBA
result\cite{Gornyi}. The two results differ only in the replacement of Bessel
functions by trigonometric ones. The discrepancy may result from
approximation (\ref{approximated_wf}) that neglected the increase
in the wave function amplitude as the classical turning points are
approached.

The extended states lie at the center of
the Landau level whereas the localized states lie at its ends. Thus, if the
chemical potential lies in the region of the localized states, the
contribution of $\Gamma_H^{(ext)}$ will show activated behavior at
low temperatures. The relative weight of the contributions of localized and extended states is
difficult to estimate, but it is possible that this activated behavior is the source of that
seen in the experiments.

The above physical understanding of $\Gamma_H$ sheds light on the non monotonous temperature
dependence of $\rho^D$. As shown in (\ref{Gamma_H_result}), $\Gamma_H$ is a weighted sum over all possible single
electron transitions.  The transition strength $T_{\beta\gamma}$, which is a
measure of the net transition rate between the two states, is
non zero only if $\epsilon_\beta = \epsilon_\gamma + \omega$. From
Eq.(\ref{sigma_D_xx}) it follows that $\omega \lesssim T$. For well separated Landau levels, $\Delta \ll \omega_c$, with
$\Delta$ being the Landau level's width. At low temperatures, $T \ll
\Delta$, the phase space for scattering increases with
temperature. Hence, $\Gamma_H$ and $\rho^D$ increase with
temperature as well. However, for $\Delta \ll T \ll \omega_c$ the
difference in occupations $n_{\mbox{\tiny F}}(\epsilon_\gamma)-n_{\mbox{\tiny F}}(\epsilon_\beta)$
scales as $1/T$ and decreases with increasing temperature. Consequently so does the net
transition rate between these states, as well as $\Gamma_H$ and $\rho^D$. Finally,
for $T \gg \omega_c$ inter Landau level transitions
become possible and $\rho^D$ returns to increase with temperature.
To fully account for the temperature dependence of $\rho^D$ the
temperature dependence of the screened inter-layer interaction,
$U$, must be incorporated as well \cite{Gornyi}. Nevertheless, the
above physical picture remains intact.

We now turn to the
derivation of Eq.(\ref{Gamma_H_result}). In terms of the exact
eigenstates $\{ \Psi_\alpha \}$ and exact energies $\{
\epsilon_\alpha \}$,
    \bea \Gamma_\perp(\bm{q},\omega) = -i
\sum_{\alpha\beta\gamma} \int \frac{d \epsilon}{2 \pi}
(v_x)_{\gamma\alpha} (\rho_{q})_{\alpha\beta} (\rho_{-q})_{\beta\gamma}  \nonumber \\
\times F(\epsilon_\alpha,\epsilon_\beta,\epsilon_\gamma,\omega),
\label{Gamma_x_qy}
    \eea
where $F$ is determined by the standard Keldysh technique\cite{Langreth, Huag,Rammer}.  The
inversion of the magnetic field leaves the energies unchanged
while transforming $\{\Psi_\alpha\}\rightarrow \{ \Psi_\alpha^* \}$. Thus,
the second term of $\Gamma_H$, given by
Eq.(\ref{Gamma_H_definition}), can be related to the first term.
As
evident from Eq.(\ref{Gamma_jDC}) the rectification function is
real. Hence
\bea
&& \Gamma_H(\bm{q},\omega) =
\sum_{\alpha\beta\gamma} \int \frac{d \epsilon}{2 \pi} {\rm Im} \{(v_x)_{\gamma
\alpha}(\rho_\bm{q})_{\alpha\beta}(\rho_{-\bm{q}})_{\beta\gamma}
 \}\nonumber \\
&& \times {\rm Re} \left\{ G_\alpha^r(\epsilon) \big(
G_\beta^r(\epsilon+\omega) - G_\beta^a(\epsilon+\omega) \big)
G_\gamma^a(\epsilon)  \right\}       \nonumber \\
&&\times \left( n_{\mbox{\tiny F}}(\epsilon) - n_{\mbox{\tiny F}}(\epsilon+\omega) \right) \ \ \ \   + \ \ \ \ \footnotesize{ \left\{
\begin{array}{cc}
\bm{q} \rightarrow -\bm{q} \\ \\
\omega \rightarrow -\omega
\end{array}
\right\} }, \label{Gamma_H_1}
\eea
where $G_\alpha^r$ and $G_\alpha^a$
are, respectively, the retarded and advanced Green functions.
After some straightforward manipulations
of which we particularly note the use of Heisenberg's equation of motion, $ (v_x)_{\gamma\alpha} = i
\left(\epsilon_{\gamma}-\epsilon_\alpha\right)(x)_{\gamma\alpha} $, we obtain
\bea
&& \Gamma_H(\bm{q},\omega) = 2 \pi \sum_{\beta \gamma} \left(
n_{\mbox{\tiny F}}(\epsilon_\gamma) - n_{\mbox{\tiny F}}(\epsilon_\beta) \right)
\delta(\epsilon_\beta - \omega - \epsilon_\gamma)   \nonumber               \\
&& {\rm Re} \left\{ (x \rho_{\bm{q}})_{\gamma \beta}
(\rho_{-\bm{q}})_{\beta\gamma} -
(x)_{\gamma\gamma}|(\rho_{\bm{q}})_{\gamma \beta}|^2   \right\} +
\footnotesize{ \left\{
\begin{array}{cc}
\bm{q} \rightarrow -\bm{q} \\
\omega \rightarrow -\omega
\end{array}
\right\} }.  \nonumber \\             \label{Gamma_H_3}
\eea
Adding the $\bm{q} \rightarrow -\bm{q}, \omega \rightarrow -\omega$ term explicitly
and using $[x,\rho_{\bm{q}}] = 0$ we obtain Eq.(\ref{Gamma_H_result}).

Eq.(\ref{Gamma_H_3}) is valid provided that $\Delta x_{\beta\gamma}$ is finite.
This is certainly true for localized states.
It follows from Eqs.(\ref{decomposition},\ref{Delta_x_localized}) that
$\Delta x_{\beta\gamma}$ is finite
for extended states as well since $(x_0)_{\gamma\gamma}-(x_0)_{\beta\beta} \approx ql_H^2$,
despite the fact that both of these expectation values are ill defined.

In summary, we studied the drag resistivity, $\rho^D$, for well
separated high Landau levels, focusing on the dependence of the
Hall rectification function $\Gamma_H$ on the chemical potential
$\mu$ and temperature $T$. Using the basis of exact single
particle eigenstates, we expressed $\Gamma_H$ as a sum over
transitions induced by the potential $V_{\bm{q}\omega}$. The
contribution of each transition is the product of its rate by the
displacement it induces.  For localized states the displacement is
$ql_H^2$, yielding a sign of $\Gamma_H$ that is independent of
$\mu$. For extended states this displacement is augmented by
another contribution, induced by the disorder potential. The sign
of the latter contribution is odd with
respect to the distance of $\mu$ from the center of the Landau
level. For high enough Landau levels this contribution dominates
and induces oscillations in $\rho_D$ as a function of density
difference between the layers. The non monotonous dependence of
$\rho^D$ on $T$ is a consequence of the strong oscillations in the
density of states of well separated Landau levels. The difference
between $\Gamma_H$ for localized and extended states may account
for the activated $T$ dependence of $\rho^D$. Our results for
$\Gamma_H$ of extended states with short range disorder are in
agreement with the SCBA results of Gornyi et al.\cite{Gornyi}, and
unravel the physical picture behind this approximation.

We are grateful to B.I. Halperin, A. Mirlin and F. von Oppen for
instructive discussions, and to the US--Israel BSF and Israel
Science Foundation for financial support.

\end{document}